\documentclass{article}

\usepackage{graphicx}  
\usepackage{amsmath}   
\usepackage{amssymb}   
\usepackage{bm}         

\usepackage[active]{srcltx}

\def\eq#1{{Eq.~(\ref{#1})}}

\def\ket#1{|#1\rangle}                    
\def\bk#1#2#3{{\langle #1|#2|#3\rangle}}  

\def\zpl{{zero-point-length}}

  \title{A class of QFTs with higher derivative field equations leading to standard dispersion relation for the particle excitations}
  \author{T. Padmanabhan\\
  IUCAA, Pune University Campus,\\
  Ganeshkhind, Pune- 411 007.\\
  {\small {email: paddy@iucaa.in}}
  }

  \date{}  

\begin{document}
  
  \maketitle
  
   \begin{abstract}
   Given any (Feynman) propagator which is Lorentz and translation invariant, it is possible to construct an action functional for a scalar field such that the quantum field theory, obtained by path integral quantization, leads to this propagator. In general, such a theory will involve derivatives of the field higher than two and can even involve derivatives of infinite order. The poles of the given propagator determine the dispersion relation for the excitations of this field. I show that it is possible to construct field theories in which the dispersion relation is the same as that of standard Klein-Gordan field, even though the Lagrangian contains derivatives of infinite order. I provide a concrete  example of this situation starting from a propagator which incorporates the effects of the \zpl\ of the spacetime. I compare the path integral approach with an alternative, operator-based approach, and highlight the advantages of using the former.
    \end{abstract}

  I begin by describing how one can construct a QFT, if we are given a Lorentz and translation invariant propagator.\footnote{One motivation for this exercise is to investigate how  the propagator with \zpl\ (see e.g., ref. \cite{pid}) can be obtained from quantizing a scalar field. But the question is of more general interest.} I will argue that this is indeed possible in the path integral formalism but there are some issues in the operator-based (canonical) quantization approach. Let us see how these results come about. 
  
  It will be useful to keep the discussion somewhat general in order to understand various nuances. Consider a scalar field theory, which, when quantized by some procedure leads to a translationally invariant 2-point function of the form: 
  \begin{equation}
  iG(x_2,x_1) \equiv iG(x_2-x_1) = \int \frac{d^4p}{(2\pi)^4} \frac{e^{-ip(x_2-x_1)}}{[F(p^2) - i \epsilon]}
    \label{piG}
   \end{equation}
  where $F(p^2)$ is a real  scalar function, in the momentum space, of its argument $p^2\equiv p_ap^a$, such that the integral exists. I want to obtain this $G(x)$, with $x\equiv (x_2-x_1)$, as the propagator for a suitable scalar quantum field theory. This is completely straightforward in the path integral approach. Consider a Lorentz invariant scalar field $\phi(x)$ with the 4-dimensional Fourier transform $\phi(p)$. I choose the action for the scalar field to be:
  \begin{equation}
   A\equiv -\frac{1}{2}\int \frac{d^4p}{(2\pi)^4} \, \phi^*(p) F[(p^2)] \phi(p) = -\frac{1}{2}\int d^4x\ \phi(x) F[(-\Box)] \phi(x)
   \label{action}
  \end{equation} 
  (In the simple case of $F(p^2) =F_{std}(p^2)\equiv -p^2+m^2$, the above action reduces to that of standard Klein-Gordon (KG) field.) It is trivial to perform a path integral quantization using the action in \eq{action}. The partition function of the theory can be defined in terms of a  path integral by 
   \begin{equation}
    Z(J) = \int \mathcal{D} \phi(x)\ \exp\left[iA[\phi(x)] + i\int d^4x\ \left[J(x) \phi (x)\right]\right]
   \end{equation} 
   I assume that $F$ in \eq{action} is replaced by $F-i\epsilon$ to ensure convergence. (This standard $i\epsilon$ regulator adds a term $i \epsilon \phi^2$ to the Lagrangian.)
   Since the action is quadratic, it is straight-forward to evaluate this Gaussian path integral and identify the propagator as the inverse of $F$.  In momentum space this will lead to the propagator function $iG(p) = [F(p^2) + i \epsilon]^{-1}$. In real space, it will lead to the propagator in \eq{piG}, which what I want. So, from the point of view of path integral quantization, it is easy to reverse-engineer any translation invariant propagator and obtain the action functional for a scalar field. 
  \textit{I did not have to worry about equal-time-commutation-relation (ETCR), canonical quantization etc. etc. thanks to the existence of the path integral.} 
    
   One can also do the same analysis in the Euclidean sector, thereby avoiding the $i\epsilon$ prescription. Here $p^2$ will be replaced by $-p_E^2$ (with $p_E^2$ being positive definite Euclidean variable) and the Euclidean action will be defined using $F(-p_E^2)$ as long as it is positive definite. The Euclidean propagator will be $G_E(p_E)=1/F(-p_E^2)$ which, of course, gives the standard result for $F_{std}(-p_E^2)=p_E^2+m^2$. The Euclidean approach is often easier to handle algebraically. 
   
   The above procedure allows you to construct a QFT which, via path integral, can produce \textit{any} translation and Lorentz invariant propagator $G(x-y)$. The simple steps in the construction are as follows: Start with a propagator $G(x-y)$ of your choice (which depends only on $(x-y)^2$ to ensure translation and Lorentz invariance),  find its Fourier transform $G(p^2)$ and define $F(p^2)$ as the  reciprocal $F(p^2)=1/G(p^2)$;  then  the action in \eq{action} with this $F(p^2)$ will give rise to the propagator $G(x-y)$ in the path integral approach.

  Let me illustrate this procedure for a propagator which incorporates the \zpl\ in flat spacetime \cite{pid}. In this case it is convenient to work in the Euclidean sector.  In the Schwinger representation of the Euclidean propagator, this propagator is given by:
   \begin{equation}
 G_{\rm QG} (x) = \frac{1}{16\pi^2}  \int_0^\infty \frac{ds}{s^2} \ \exp\left(-sm^2 - \frac{x^2+L^2}{4s}\right)
 =\frac{1}{4\pi^2}\frac{m}{\sqrt{x^2+L^2}}K_1[m\sqrt{x^2+L^2}]
  \label{X1}
 \end{equation} 
 This involves replacing $x^2$ by $x^2+L^2$ in the Schwinger kernel.\footnote{There is an extensive discussion of this propagator in the literature; some of the earlier papers are \cite{pid,zplextra} and for a sample of more recent work, see \cite{zplrecent}. In this work, I use $G_{\rm QG} (x)$ purely as a non-trivial example to illustrate the ideas developed here, without going into details of how $G_{\rm QG} (x)$ is obtained etc. Those who are interested in more details of $G_{QG}$ (like e.g., the motivation, construction, properties etc.), can find these in the cited references; in particular, some of these aspects are succinctly summarized in ref. \cite{tpnew}.}
 The momentum space propagator is obtained by a Fourier transform: 
 \begin{equation}
  G_{\rm QG}(p^2) = \int_0^\infty ds  \exp\left[-s(p^2+m^2) - \frac{L^2}{4s}\right] = \frac{L}{\sqrt{p^2+m^2}} \ K_1[L\sqrt{p^2+m^2}]
   \label{start1}
  \end{equation} 
 Clearly, we get back the standard results (like e.g. $G_{std}(p^2)=(p^2+m^2)^{-1}$) when $L=0$, as we should. Since $K_1(x)$ is a monotonically decreasing, positive function of its argument with $K_1(x)\approx 1/x$ near $x=0$, it is clear that the only pole  $G_{\rm QG}(p^2)$ occurs at
 $p^2=-m^2$, with unit residue. (More precisely, on analytic continuation to the Lorentzian sector with $p_E^2\to -p^2$, the pole is at $p^2=m^2$ with unit residue.) To construct a QFT which will lead to the propagator $ G_{\rm QG}(p^2)$, we identify the function $F$ in \eq{action} as:
 \begin{equation}
  F_{QG}(p^2)=\frac{1}{G_{\rm QG}(p^2)}=\frac{\sqrt{p^2+m^2}}{LK_1[L\sqrt{p^2+m^2}]}
  \label{Fzpl}
 \end{equation} 
 The QFT is now defined in terms of the action in \eq{action} with this choice of  $F_{QG}(p^2)$ in the momentum space. The real space version in the \textit{in the Lorentzian sector} is obtained by the standard replacements $p^2_E\to -p^2\to \square$, and the addition of $-i\epsilon$ regulator, leading to:
 \begin{equation}
  F_{QG}(\square)=\frac{\sqrt{\square+m^2}}{LK_1[L\sqrt{\square+m^2}]}-i\epsilon \qquad {\rm (Lorentzian\; sector)}
  \label{Fzpl1}
 \end{equation}
 Obviously, the theory reduces to the Klein-Gordon limit, with $F_{std}(\square)=\square+m^2-i\epsilon $, in the limit of $L\to0$. So we have succeeded in obtaining our propagator via path integral quantization of a scalar field with a non-trivial action functional.

 As an aside let me comment on the heat kernel for the theory. We have the trivial identity:
    \begin{equation}
    G(p^2) = \int_0^\infty ds \ e^{-s F(p^2)}; \qquad F(p^2)=1/G(p^2)
   \label{one}
  \end{equation} 
  which, of course holds for any function $G(p^2)$.   The  propagator in space(time),   given by the Fourier transform of $G(p^2)$, can be expressed in the Schwinger proper time representation as:
  \begin{equation}
  G(x^2) = \int \frac{d^4p}{(2\pi)^4} \, e^{ip\cdot (x_2 - x_2)} G(p^2) 
  = \int_0^\infty ds \bk{x_2}{e^{-sF(-\Box)}}{x_1}
  \equiv \int_0^\infty ds\ \mathcal{H}(x_2,x_1;s)
   \label{two}
  \end{equation} 
  where $x=x_2-x_1$ and the last equality defines the integrand $\mathcal{H}(x_2,x_1;s)$. It is easy to verify that $\mathcal{H}$ satisfies the standard heat kernel equation and the boundary condition,
  \begin{equation}
  \frac{\partial\mathcal{H}}{\partial s} + F(-\Box)\mathcal{H} =0\, ; \qquad \mathcal{H}(x_2,x_1;0) = \delta (x_2 - x_1)
   \label{three}
  \end{equation} 
  So $\mathcal{H}(x_2,x_1;s)$ can indeed be thought of as the heat kernel for the QFT we have constructed. In momentum space, this heat kernel is given by:
  \begin{equation}
  \mathcal{H}(p;s)=\exp-sF(p^2) =\exp -\left[ \frac{s\sqrt{p^2+m^2}}{LK_1[L\sqrt{p^2+m^2}]}         \right]
  \label{barH0}
  \end{equation} 
  On the other hand, we can read off from \eq{start1} \textit{another} function $\bar{\mathcal{H}}$ by writing:
  \begin{equation}
  G_{\rm QG}(p^2) = \int_0^\infty ds  \exp\left[-s(p^2+m^2) - \frac{L^2}{4s}\right]  \equiv \int_0^\infty ds\ \bar{\mathcal{H}}(p;s);
  \label{start1new}
  \end{equation} 
  with
  \begin{equation}
   \bar{\mathcal{H}}(p;s)=\exp\left[-s(p^2+m^2) - \frac{L^2}{4s}\right]
   \label{barH}
  \end{equation} 
 Both  $\bar{\mathcal{H}}(p;s)$ (in \eq{barH}) and $\mathcal{H}(p;s)$ (in \eq{barH0}) lead to the same propagator on integration over $s$. Of these two,  $\mathcal{H}$ satisfies the standard heat kernel equation and the boundary condition while  $\bar{\mathcal{H}}$ does not. So while the propagator is unique, its Schwinger propertime representation is not.
   
 Let us now consider the excitations and their dispersion relation in such quantum field theories. In general, the poles of the propagator $G(p^2)$, or --- equivalently ---
 the zeros of $F(p^2)$ determine the dispersion relation. In the simplest case of Klein-Gordan theory, with $F_{std}(p^2)=-p^2+m^2$, the dispersion relation is given by the standard one $p^2=m^2$. Of course, for a general $F(p^2)$ the interpretation will not be easy. For example, if 
 $F(p^2)=(-p^2+m^2) (-p^2-M^2)$ we will have a tachyonic state with $p^2=-M^2$ as well in the model; if the poles are at complex values one has to describe them in terms of instabilities, decay  etc.
 
 Remarkably enough, there is a class of theories for which these issues do \textit{not} arise and the dispersion relation is still given by the standard one, $p^2=m^2$ \textit{even though the propagator is non-trivially different} from the KG propagator. To identify this class of theories, 
  I will again start with the propagator defined through \eq{piG} but now impose  two conditions on the function $F(p^2)$:
 
 \medskip
 (a) The only  solution to $F(p^2) =0$ --- in the entire complex plane --- is $p^2=m^2$ where $m$ is a real, Lorentz invariant, parameter with the dimensions of inverse length.
 
 \medskip
 (b) $|F'(m^2)| = 1$. 
 \medskip
 
 \noindent
 This is same as demanding that the function $1/F(p^2)$ has \textit{only one} pole in the complex plane at $p^2=m^2$ and the residue at the pole is unity. Obviously, the choice $F_{std}(p^2)=-p^2+m^2$ will satisfy both these conditions and will lead to the standard QFT of a massive scalar field. It is also easy to see that the function $F_{QG}(p^2)$ in \eq{Fzpl},  for example,  also satisfies these conditions. If two different functions --- viz., $1/F_{QG}(p^2)$ and $1/F_{std}=1/(-p^2+m^2)$ --- have identical poles and residues in the complex plane, then their difference will be an entire function $f(z)$. \textit{If $f(z)$ is bounded}, then it has to be a constant. However, when $f(z)$ is unbounded, we will get a theory which is non trivially different from the standard QFT.  This is precisely which happens in the case of the function $F_{QG}(p^2)$ in \eq{Fzpl}. Also note that, while the integral in \eq{piG} exists for the choice of $F_{QG}(p^2)$ in \eq{Fzpl}, it cannot be evaluated by the usual procedure of closing the complex contour and picking up the residues; this is because $F_{QG}(p^2)$ in \eq{Fzpl} differs from $(-p^2+m^2-i\epsilon)$ by an unbounded function in the complex plane.
 
 To summarize, QFTs with the action in \eq{action} with $F(p^2)$ satisfying the conditions (a) and (b) above have particle excitations which obey the standard KG dispersion relation. But these theories are structurally very different from the KG scalar field theory and have a non-trivially different propagator $1/F(p^2)$. A simple, physically relevant, example is given by the propagator with \zpl\ and the associated QFT with $F(p^2)$ in \eq{Fzpl}. As a bonus, we also see that the addition of the \zpl\ by this procedure does not modify the dispersion relation for the excitations.

 Let us now ask what happens when we try to quantize the field $\phi$ in the action in \eq{action} by the usual canonical quantization procedure. 
 The standard approach to canonical quantization involves identifying the canonical momentum $\pi$ conjugate to the field $\phi$ and imposing the ETCR on $\phi$ and $\pi$.  When $F(q)$ is not a linear function of $q$, i.e., when we go beyond the Klein-Gordon structure, you cannot write the action in \eq{action} with a quadratic kinetic energy term. This implies that the momentum conjugate to $\phi$ is no longer given by $\partial_t \phi$.  Therefore the standard procedure of imposing ETCR and  then deriving the commutators for  creation and  an annihilation operators etc. will not work. 
 
 When $F(p^2)$ is a polynomial (like e.g, a product of factors $(-p^2+m_i^2)$ with $i=1,2,...N$) there is another standard procedure to perform the canonical quantization  by introducing a set of auxiliary fields, essentially to reduce the degree of derivatives appearing in the Lagrangian (see e.g., \cite{gwg} and references therein.). In this case, $G(p^2)$ will have $N$ poles at $m_i^2$ with $i=1,2,...N$. But the case I am interested in has a non-polynomial $F(p^2)$ and --- more crucially --- the $G(p^2)$ has \textit{only one} pole at $p^2=m^2$; see e.g the $F(p^2)$ in \eq{Fzpl}. \textit{This means the dispersion relation for excitations in the theory is the same as that of the standard KG field.} If one  attempts a canonical quantization with auxiliary fields, we will now need \textit{infinite} number of auxiliary fields. It is not clear how they will all combine together to give the standard dispersion relation for the excitations of the theory.

There is another way to approach this issue, taking advantage of the fact $F(p^2)$ satisfies the two conditions (a) and (b) mentioned earlier.
 I will call this procedure canonical-\textit{like} quantization since there are some key differences.
  The action will now involve --- for non-polynomial $F$, like the one in \eq{Fzpl} we are interested in ---   infinite number of derivative terms if $F$ is defined through a Taylor series expansion.
 The variation of the action in \eq{action}, in the momentum space,  leads to the classical equations of motion for the theory given by $ F(p^2) \phi(p) =0$.  This tells you that the solution $\phi(p) $ should have the structure 
  \begin{equation}
   \phi(p)= \delta_D [F(p^2)]\ Q(p)
  \end{equation} 
  where $Q(p) $ is an arbitrary function which is regular at $p^2=m^2$. The conditions (a) and (b) on the function $F(p^2)$ allow us to simplify the Dirac delta function  $\delta_D [F(p^2)]$. Using the standard rule for the Dirac delta function of the argument and the conditions (a) and (b) satisfied by the function $F$ we immediately see that the contribution from the \textit{only} root gives $\delta_D [F(p^2)] = \delta_D(-p^2+m^2)$. Therefore we can write the  solutions of the field equation as 
  \begin{equation}
     \phi(p)= \delta_D [F(p^2)] Q(p) = \delta_D(-p^2+m^2)Q(p)
  \end{equation} 
  \textit{This tells you that the solutions to the field equation have the same on-shell structure as the solutions to the standard Klein-Gordon equation which we obtain when} $F=F_{\rm std} =-p^2 + m^2$. So, a general solution to the real scalar field can be expressed as a linear superposition 
  \begin{equation}
   \phi(x) = \int \frac{d^3{\bm p}}{(2\pi)^3} \frac{1}{2\omega_{\bm p}} \left( A_{\bm p} e^{-ipx} + c.c\right);\quad \omega_{\bm p}=\sqrt{\bm{p}^2+m^2}
   \label{expansion}
  \end{equation} 
  where $px \equiv \omega_{\bm p} t - \bm{p\cdot x}$ and $A_{\bm p} $ is an arbitrary function. The structure on the right hand side is Lorentz invariant and in fact is identical to what you start with in the case of Klein-Gordon equation.\footnote{
    In arriving at this result, I used the fact that the \textit{only} zero of $F(p^2)$  occurs in the real line. Any  zero off the real line  will, in general lead to mode functions (solutions) which are unbounded rather than oscillatory. So if we insist that the modes of the scalar field should not exhibit any instability, then the two conditions (a) and (b) which we imposed  on $F(p^2)$ are mandatory.}
  
  Let us now try to give a quantum meaning to the classical solution in \eq{expansion}. 
  In the absence of useful canonical momentum and ETCR, we will
  use  an alternative procedure: We  start by promoting the expansion in \eq{expansion} directly into an operator-valued equation and then \textit{postulate} the commutation rule 
  \begin{equation}
   [A_{\bm p}, A^\dagger_{\bm q}] =  (2\omega_{\bm p}) \, (2\pi)^3\delta_D(\bm{p-q})
   \label{comm1}
  \end{equation}
  This uses the Lorentz invariant structure for Dirac delta function; usually one works with the rescaled operators $a_{\bm p} \equiv \sqrt{2\omega_{\bm p}} \, A_{\bm p}$ so that $ [a_{\bm p}, a^\dagger_{\bm q}] =  (2\pi)^3 \delta_D(\bm{p-q})$. We can then introduce the number operator $n_{\bm k} \equiv  a^\dagger_{\bm k} a_{\bm k}$ with integer-labeled eigenkets and identify a Fock basis of kets $\ket{\{n_{\bm k}\}}$. That is, instead of identifying a canonical momentum from the Lagrangian and imposing ETCR, we are directly promoting the \textit{solution} to the field equation, given  in \eq{expansion} to operator-valued equation by introducing creation, annihilation operators and the Fock basis. This, in turn, will imply, purely algebraically, that the commutator of $\phi(t,\bm{x})$ and $\partial_t\phi (t, \bm{y})$  happens to be $\delta_D(\bm{x-y})$; however, this is now just a consequence of our postulate in \eq{comm1} and we do \textit{not} identify $\partial_t \phi$ with some canonically conjugate momentum or impose ETCR between field and canonically conjugate momentum. 
  
  So far, I have not introduced a Hamiltonian for the theory. This is, however, possible along the following lines. We already know the field equation and its solution  [given by \eq{expansion}]. So we already know the \textit{complete} time evolution of the operator $\phi(t,\bm{x})$ or --- equivalently --- the time evolution of creation and annihilation operators. From \eq{expansion} we see that we can think of creation and annihilation operators having the following time evolution:
  \begin{equation}
    a_{\bm k}(t) = a_{\bm k}(0) \ e^{-i \omega_{\bm k} t}; \qquad  a_{\bm k}^\dagger(t) = a_{\bm k}^\dagger(0) \ e^{i \omega_{\bm k} t}
  \end{equation} 
  with the identification of $a_{\bm k}$ appearing in the expansion \eq{expansion} as  $a_{\bm k}(0)$.
 All I need to do is to identify a Hamiltonian operator which will induce this time evolution by standard Heisenberg equation of motion $i\partial_t a_{\bm k}(t) = [a_{\bm k},H]$. (This will ensure that  $i\partial_t\phi=[\phi,H]$). This is easy because
    we know from standard analysis that, this will indeed be true if I postulate the Hamiltonian to be the sum:
  \begin{equation}
  H =  \int\frac{d^3{\bm k}}{(2\pi)^3} H_{\bm k} =  \int \frac{d^3{\bm k}}{(2\pi)^3}\omega_{\bm k} a_{\bm k}^\dagger a_{\bm k}
   \label{hamil}
  \end{equation} 
  It immediately follows that the energy of the state $\ket{\{n_{\bm k}\}}$ is given by the sum of $E_{\bm k} = \omega_{\bm k} n_{\bm k}$. So the physical states, the excitations spectra and the time evolution remains identical to the standard Klein-Gordon theory. 
  Let me summarize the logical structure of what I have done so far, vis-a-vis canonical-like quantization: 
  
  (i) I start with a well-defined \textit{classical} field theory, generically involving derivatives higher that second order, based on a function $F$ which satisfies two conditions (a) and (b). From the action functional in \eq{action} I obtain the field equation and a solution given in \eq{expansion}. 
  
  (ii) To quantize the theory, I upgrade \eq{expansion} as an operator equation and \textit{postulate} the commutation rules in \eq{comm1}. This replaces the \textit{postulate} of ETCR in the standard approach.
  
  (iii)  Since I already know the time evolution of the field operator $\phi$, I determine the Hamiltonian from the condition that standard Heisenberg equations of motion, $i\partial_t\phi=[\phi,H]$ should hold. This gives me the Hamiltonian for the theory in \eq{hamil}. 
  
  (iv) I do not use the classical action function to identify the conjugate momentum or to derive the form of the Hamiltonian. 
  The conjugate momentum is needed only for ETCR and I replace the postulate of ETCR by the postulate in \eq{comm1}. The Hamiltonian is then derived/defined such that  the time evolution is consistent with the operator-valued field equation.
  
  These results  should be  --- at least mildly --- surprising. They show that one can have a very general field equation (with an arbitrary function $F$) for which the excitation spectrum  is the \textit{same} as standard Klein-Gordon equation, representing spinless bosonic particles obeying the relativistic relation between energy and momentum. More importantly, \textit{this canonical-like quantization procedure,  will lead to a completely different propagator} compared to what we found in the case of path integral. This arises as follows: We now have a well defined vacuum state and field operator; so one could define the propagator as the expectation value of the time ordered product of the field operators. This will lead, on using \eq{expansion}, to the standard result:
   \begin{equation}
   G (x_2,x_1) = \bk{0}{T[\phi(x_2)\phi(x_1)]}{0} = \int    \frac{d^3 \bm{k}}{(2\pi)^3}\ \frac{1}{2\omega_{\bm k}}        \ e^{-i\omega_{\bm k}|t| + i \bm{k}\cdot \bm{x}} ; \qquad x = x_2 - x_1
  \end{equation}
  Incidentally, if one determines the dispersion relation of the theory by looking at the poles of the propagator, we get the same, standard dispersion relation for all propagators of the form in \eq{piG}, because of our assumptions (a) and (b) about the function $F(p^2)$. So even though the propagators obtained by the two methods are widely different, the dispersion relation obtained from their poles remain the same.
  
     It is rather interesting that the path integral actually picks up the propagator we started with, in \eq{piG}, at one go while it is unclear how to obtain it by canonical quantization procedure, after introducing infinite number of auxiliary fields. If we take that approach we will have infinite pairs of propagators which, somehow, should combine together to give \eq{piG}. This connection is worth investigating.\footnote{More generally, it is not clear whether this approach --- canonical-like quantisation, with a rather ad hoc identification of Hamiltonian etc. --- leads to a completely consistent quantum theory. This requires further study.} 

\medskip     
\noindent To summarize, the key results of this work are the following:  
\begin{itemize}
 \item 
 Given any Lorentz and translation invariant propagator in \eq{piG}, it is always possible to construct a scalar field action [see \eq{action}] such that path integral quantization of the theory leads to the propagator in \eq{piG}. A non-trivial example is given by the propagator with \zpl\ (see \eq{start1}), which can be obtained from a scalar field theory with the choice of operator in \eq{Fzpl1}.
 \item
 For a general propagator of the form in \eq{piG}, the excitation spectra and the dispersion relation can be complicated and even ill-defined. However, there exists a subset of these models in which $F(p^2)$ satisfies two conditions [listed as (a) and (b) above] for which a simple situation arises; in these models, the excitation spectra and the dispersion relation are identical to those in standard KG theory. 
 \item
 The propagator with \zpl\ belongs to this subset of models. So we conclude that the addition of \zpl\ does not change either the nature of particle excitations or modify the dispersion relations of the theory.
 \item
 It is also possible to attempt an operator-based quantization of the action in \eq{action} when $F(p^2)$ satisfies two conditions [listed as (a) and (b) above]. This leads to a structure very similar to that of KG theory as far as the Fock basis is concerned. The propagator, defined as time ordered correlator, however, will not match with the one obtained from the path integral quantization; nevertheless, the dispersion relations are identical.

\end{itemize}

 \section*{Acknowledgement}

I thank S. Date, A.D. Patel and S. Solodukhin for useful discussions.  My research  is partially supported by the J.C.Bose Fellowship of Department of Science and Technology, Government of India.


\begin{thebibliography}{000}

\bibitem{pid}

 T. Padmanabhan,  \textit{Phys. Rev. Letts}, \textbf{78}, 1854 (1997) [hep-th-9608182]; 
  
  T. Padmanabhan,  \textit{Phys. Rev.}, \textbf{D 57} , 6206 (1998) 
  
 \bibitem{zplextra} 
 B. S. DeWitt, (1964) \textit{Phys. Rev. Lett.} \textbf{13}, 114; 
  
  T.Padmanabhan, (1985), \textit{Gen. Rel. Grav.,}  \textbf{17}, 215; 
  
  T.Padmanabhan, (1985), \textit{Ann. Phys.},  \textbf{165}, 38; 
  
  T.Padmanabhan, (1987), \textit{Class. Quan. Grav.},  \textbf{4}, L107;
 
  S. Shankaranarayanan and T. Padmanabhan, \textit{Int. Jour. Mod. Phys }, \textbf{D 10}, 351 (2001) [gr-qc-0003058];

K.Srinivasan, L.Sriramkumar and T. Padmanabhan,  \textit{Phys. Rev.} \textbf{D 58}, 044009 (1998) [gr-qc-9710104]

Michele Fontanini, Euro Spallucci, T. Padmanabhan, \textit{Phys.Lett.,} \textbf{B633}, 627-630 (2006) [hep-th/0509090]  
  

 
 \bibitem{zplrecent}
 
Dawood Kothawala, L. Sriramkumar, S. Shankaranarayanan, T. Padmanabhan,  \textit{Phys.Rev.,} \textbf{D 79}, 104020 (2009) [arXiv:0904.3217]

  Kothawala D and Padmanabhan T (2014), \textit{Phys. Rev.} \textbf{D 90} 124060 [arXiv:1405.4967];
  
  Kothawala D (2013) \textit{Phys. Rev.} \textbf{D 88} 104029;
  
  Padmanabhan, T (2015) \textit{Entropy}, \textbf{17},  7420 [arXiv:1508.06286];
  
  D. Jaffino Stargen, D. Kothawala, (2015), \textit{Phys.Rev.} \textbf{D 92}  024046 [arXiv:1503.03793]
  
  Nahomi Kan et al., [arXiv:2007.00220];   [arXiv:2004.07527];
  
  Erik Curiel, Felix Finster, J.M. Isidro, \textit{Symmetry} \textbf{12}(1) (2020) 138;
 
\bibitem{tpnew}

T.Padmanabhan, \textit{Phys. Letts.},\textbf{B 809} (2020) 135774 

T. Padmanabhan,  \textit{J. High Energ. Phys.}, 2020, \textbf{13} (2020) [arXiv:2006.06701].

 \bibitem{gwg}
 
 See e.g., G.W. Gibbons et al., \textit{Phys. Rev.} \textbf{D 100}, 105008 (2019) [arXiv:1907.03791]
  
 \end{thebibliography}
 \end{document}